\documentclass[a4paper, 10pt, conference]{IEEEtran}
\IEEEoverridecommandlockouts

\usepackage{graphics}
\usepackage{graphicx}
\usepackage{epsfig,amssymb}
\usepackage{amsmath}
\usepackage{times}
\usepackage{mathrsfs}
\usepackage{array}
\usepackage{amsmath, latexsym, amsfonts, amssymb}
\usepackage[mathscr]{eucal}
\usepackage{array, tabularx}
\usepackage[table]{xcolor}
\usepackage{psfrag}
\usepackage{multirow}
\usepackage{booktabs}
\usepackage{epsfig}
\usepackage{cite}
\usepackage{tikz}
\usetikzlibrary{shapes.geometric, arrows}
\usepackage{algorithmic}
\usepackage{algorithm}
\usepackage{ amssymb }
\usepackage{threeparttable}

\usepackage{subfigure}
\usepackage{url}

\newcommand{\paren}[1]{\left(#1\right)}
\newcommand{\sqparen}[1]{\left[#1\right]}
\newcommand{\brparen}[1]{\left\{#1\right\}}

\newcommand{\I}[1]{\ensuremath{\mathsf{1}_{\left\{#1\right\}}}} 
\newcommand{\PR}[1]{\ensuremath{\mathsf{Pr}\left\{#1\right\}}} 
\newcommand{\ES}[1]{\ensuremath{\mathsf{E}\left[#1 \right]}} 
\newcommand{\e}[1]{\ensuremath{{\rm e}^{#1}}} 

\newcommand{\sinr}{\ensuremath{{\rm SINR}}}

\renewcommand{\vec}[1]{\ensuremath{\boldsymbol{#1}}} 
\newcommand{\ie}{\ensuremath{{\text{\em i.e.}}}}
\newcommand{\slrp}{\ensuremath{{\rm SLP}}}
\newcommand{\aslrp}{\ensuremath{{\rm ASLP}}}
\newcommand{\spp}{\ensuremath{{\rm SP}}}

\begin{document}

\title{A Novel Content Caching and Delivery Scheme for Millimeter Wave Device-to-Device Communications }
\author{
\IEEEauthorblockN{Theshani Nuradha\IEEEauthorrefmark{1}, Tharaka Samarasinghe\IEEEauthorrefmark{1}\IEEEauthorrefmark{2},  Kasun T. Hemachandra\IEEEauthorrefmark{1}}
\IEEEauthorblockA{
\IEEEauthorrefmark{1}Department of Electronic and Telecommunication Engineering, University of Moratuwa, Moratuwa, Sri Lanka \\
\IEEEauthorrefmark{2}Department of Electrical and Electronic Engineering, University of Melbourne, Victoria, Australia \\
\IEEEauthorblockA{
Email: theshanin@uom.lk, tharakas@uom.lk, kasunh@uom.lk
}
}
\thanks{
This work is supported by the Senate Research Council, University of Moratuwa, Sri Lanka, under grant SRC/LT/2018/2.
 } 
 \vspace{-1 cm}
}

\maketitle

\begin{abstract}
A novel content caching strategy is proposed for a cache enabled device-to-device (D2D) network where the user devices are allowed to communicate using millimeter wave (mmWave) D2D links ($>$ 6 GHz) as well as conventional sub 6 GHz cellular links. The proposed content placement strategy maximizes the successful content delivery probability of a line of sight D2D link. Furthermore, a heuristic algorithm is proposed for efficient content delivery. The overall scheme improves the successful traffic offloading gain of the network compared to conventional cache-hit maximizing content placement and delivery strategies. Significant energy efficiency improvements can also be achieved in ultra-dense networks. 
\end{abstract}

\IEEEpeerreviewmaketitle
\section{Introduction}
Small cells and device-to-device (D2D) communications are envisioned to be promising technologies for enhancing the quality of service (QoS), throughput and energy efficiency of next generation wireless networks \cite{D2D_survey2018, 5GD2D_2018}. Due to scarcity in existing cellular spectrum, millimeter wave (mmWave) frequencies have been considered 
as an enabling technology for high speed D2D communications. In D2D aided cellular networks, the successful establishment of D2D connections depends on the availability of popular files in proximity devices. Therefore, content placement in user devices is of paramount importance for D2D based traffic offloading. This paper presents a novel content caching strategy for a cache enabled D2D network, where the user devices are allowed to communicate using mmWave D2D links as well as conventional sub 6 GHz cellular links.

Cache placement schemes based on cache hit probability maximization \cite{whereToCache}, and cache aided throughput maximization \cite{hitVSthroughput}, can be found in the literature for conventional cellular networks. In these works, optimal caching probabilities are obtained such that the achieved content diversity leads to better network performance.
When it comes to mmWave networks, a D2D aware caching policy that splits the most popular content into two content groups, and randomly distributes the content groups among the users is proposed in \cite{awareDeviceCaching}. The partitioning of the most popular content is performed based on fairness considerations, such that the two content groups have equal self cache hit probability in a user device. The content placement does not consider the characteristics of mmWave propagation and the effects of blockage when placing content. A cache hit probability maximization based optimal cache placement in a mmWave ad-hoc network is studied in \cite{adhoc-mmwave}. The paper omits the effect of interference from other D2D links, which can be crucial factor on network performance.


In this paper, we consider a cloud radio access network (C-RAN) operating in the sub 6 GHz band, and supports D2D communications using mmWave spectrum. A content placement scheme is proposed for user device caching, considering the propagation characteristics of mmWave links and the interference from other D2D links, which makes it different to \cite{awareDeviceCaching} and \cite{adhoc-mmwave}. In addition, it also considers the popularity and application specific QoS constraints of different files, which makes it more applicable to next generation wireless networks, where QoS measures such as latency are considered to be key performance indicators. Thus, we maximize a different metric, referred to as the successful content reception probability, which is the probability of reception without violating the file specific QoS constraints.    
The main contributions of the paper are as follows: 
\begin{itemize}
    \item A content placement scheme is proposed for user devices by solving an optimization problem, that maximizes the successful content delivery probability within the line of sight (LoS) region of a D2D transmitter. Optimal caching probabilities of a multitude of heterogeneous files, that have their own rate constraints for successful reception, are obtained.
    \item The cache placement scheme is coupled with a user association scheme to further improve the offloading gain without violating QoS constraints.
    \item The performance of the network in terms of 
    successful delivery of content and offloading gain
   is evaluated both analytically and through simulations, to clearly highlight the gains of the proposed content placement and the user association schemes with respective to energy efficiency as well. 
\end{itemize}
The overall scheme improves the successful traffic offloading gain of the network compared to conventional cache-hit maximizing content placement and delivery strategies. Significant energy efficiency improvements can also be achieved in ultra-dense networks.


The paper organization is as follows. Section \ref{Section:System model} presents the system model and the problem formulation. The solution to the optimization problem which leads to the content placement strategy, and the user association scheme are presented in Section \ref{Section:Section System Design}. The performance of the proposed schemes are evaluated theoretically and numerically in Section \ref{section: Performance Analysis} and Section \ref{section: Numerical results}, respectively. Section \ref{section: conclusion} concludes the paper.

\section{System Model And Problem Formulation }\label{Section:System model}
\subsection{Topological Model}
We consider a C-RAN, where the remote radio heads (RRHs) are spatially distributed according to a homogeneous Poisson point process (HPPP) $\Phi_{R}$ with intensity $\lambda_{R}$. The RRHs are wirelessly connected to the edge cloud (fronthaul links) and the edge clouds are connected to the core network (backhaul links). The spatial distribution of the mobile user (MU) devices is  modelled using an independent HPPP $\Phi_{u}$ with intensity $\lambda_{u}$. Content to a MU may either be delivered from a RRH through a cellular link that operates in the sub 6 GHz band (transmit power $P_{c}$, wavelength $W_{c}$, bandwidth $B_{c}$) or from another MU through a mmWave D2D link (transmit power $P_{d}$, wavelength $W_{d}$, bandwidth $B_{d}$). Content delivery through a D2D link will only be possible if the requested file is in the cache of another proximity MU. Each MU is capable of caching $M_{d}$ files of equal size. 


RRHs are equipped with omni-directional transmitting antennas while the MUs are equipped with omni-directional antennas for cellular communications and directional antennas for mmWave communications. Similar to \cite{awareDeviceCaching}, sectorized antenna pattern is adopted at the transmitters to approximate the antenna pattern for the mmWave links. 
It is assumed that the antennas of the transmitter and the receiver are perfectly aligned for desired links while the interfering transmitter antenna bore-sight is uniformly distributed over $[0,2\pi]$. This means the probability distribution of the i.i.d. random antenna gain $G^{\prime}$ associated an interfering mmWave link is given by
$
\PR{ G^{\prime} = G_{m}^{2}}= \paren{\frac{\Delta\theta}{2\pi}}^2 ,
$
$
\PR{ G^{\prime} = G_{s}^{2}}=\paren{\frac{ 2\pi-\Delta\theta}{2\pi}}^2,
$
and
$
    \PR{ G^{\prime} = G_{m}G_{s}}= \frac{2\Delta\theta \paren{2\pi-\Delta\theta}}{\paren{2\pi}^2},
$
where $G_{m}$ and $G_{s}$ denote the main and side lobe gains, respectively, and $\Delta\theta$ denotes the angle of deviation from the antenna bore-sight. 

We refer to the MUs that request content as active MUs. At a given time, the probability of an MU being active is $\rho$. The remaining MUs, which we refer to as inactive MUs, can serve as potential D2D transmitters. It is assumed that all RRHs are active at a given time. Without loss of generality, we consider a typical MU located at the origin for our analysis. 
\vspace{-1mm}
\subsection{Channel Model} \label{SysMod}
\vspace{-1mm}
For the cellular links that operate below 6 GHz, the simple path loss model with a path loss exponent $\alpha_{c}$ is used to model the location dependent path loss. 
For mmWave links, the average LoS ball model is used \cite{coverageRateLosBall,losBall}. According to this model, a link is considered as a LoS link if the link is shorter than $D_{L}$. Otherwise, it is considered a non-line of sight (NLoS) link. The average size and the density of the blockages determine $D_{L}$ \cite{journalmmwave}. For mmWave links, blockage effects induce different path loss exponents $\alpha_{L}$ and $\alpha_{N}$ for LoS and NLoS links, respectively, with $\alpha_{L}<\alpha_{N}$. 
We assume fast Rayleigh fading where the fading power is exponentially distributed with unit mean \cite{awareDeviceCaching}. 
The received signal-to-interference-plus-noise-ratio ($\sinr$) when 
receiving file $i$ from the D2D transmitter located at $x \in \Phi_{d}$ is given by 
\begin{equation}\label{SINRcomplete}
\sinr_{d,i,x}=\frac{G_{m}^{2}h_{x} r_{x}^{-\alpha_{d}}}{ \hat{N}+\sum_{y \in \Phi_{d} \setminus \brparen{{x}}}h_{y} G^{\prime}r_{y}^{-\alpha_{d}}} ,
\end{equation}
where $\Phi_{d}$, $h_x$ and $r_x$ denote the point process of the active D2D transmitters, fading power and the distance between the MU and the D2D transmitter at $x$, respectively, $\alpha_{d} \in \brparen{\alpha_{L},\alpha_{N}}$, $\hat{N}=\frac{16 \pi^2N_{o}F_{N}B_{d}}{P_{d}W_{d}^2}$, $N_{o}$ is the noise power spectral density and $F_{N}$ is the noise figure of the receiver. Similarly, an expression for the received
$\sinr$ when receiving file $i$ through a cellular link from the RRH $x \in \Phi_{R}$, which we denote by $\sinr_{c,i,x}$, can be obtained by replacing subscript $d$ in (\ref{SINRcomplete}) with subscript $c$ and by replacing both antenna gains $G_{m}^2$ and $G^{\prime}$ by $G_{T}G_{R}$, where $G_{T}$ and $G_{R}$ are the antenna gains of the transmitting RRH and the receiving MU, respectively. 

\subsection{Content Placement}
We assume that MUs request content from a finite content library of $N$ files of equal size, and the file requests follow a Zipf distribution of popularity exponent $\epsilon$. Thus, the probability of requesting the $i$-th most popular file is given by 
$$
\beta_{i} = \frac{i^{-\epsilon}}{\sum_{j=1}^{N} j^{-\epsilon}}.
$$
The rate and the delay constraints of the files may vary with the file type and the associated application. The rate constraint for the $i$-th most popular file is denoted by $R_{i}$, and this constraint necessitates an $\sinr$ greater than
$
T_{i} =2^{ \frac{R_{i}}{B} }-1 ,
$
for a link having a bandwidth of $B \in $ $ \brparen{B_{d},B_{c}}$.

 Due to the limited storage capacity of the MUs, the propagation characteristics of mmWaves, and the rate requirements of different files, caching content at MUs should be done in an efficient manner. In this paper, we focus on designing a content placement scheme that offloads the traffic to the D2D devices without violating the rate (QoS) constraints, which are considered to be crucial in next generation networks. Moreover, considering the effects of blockage, it is preferred to have LoS D2D links. Thus, we focus on the successful LoS reception probability ($\slrp$) for the $i$-th most popular file, which we define as
 \begin{equation}\label{scdpLOS}
  \slrp_{i} =\PR{\sinr_{d,i,\hat{x}} \geq T_{i} \cap  r_{\hat{x}} \leq D_{L}},
\end{equation} where  $\hat{x}$ denotes the location of the closest D2D transmitter who has the $i$-th file in its cache. The $\slrp$ depicts the probability of receiving content from the nearest LoS D2D transmitter without violating the rate threshold.  
 
Let the  probability of the $i$-th file being stored in the cache of an MU be $q_{i}$. We define $\vec{q}$, an $N$-dimensional caching probability vector $\vec{q} = [q_{1},..,q_{N}]$, which denotes the probabilities of an MU caching the $N$ files in the content library. 
We focus on finding $\vec{q}$ that maximizes the average successful LoS reception probability ($\aslrp$), defined as $$\aslrp(\vec{q})=\sum_{i=1}^{N} \beta_{i} \slrp_{i},$$ where the $\slrp_i$ values are averaged over the request probabilities, while not violating the MU storage constraints statistically (on average). Note that our objective function captures both popularity and the QoS requirements of different content, and also the effects of wireless propagation. The optimization problem can be formulated as 
\begin{equation} \label{optimization}
\begin{aligned}
& \underset{\vec{q}}{\text{maximize}}
& &\aslrp(\vec{q})=\sum_{i=1}^{N} \beta_{i} \slrp_{i} \\
& \text{subject to}
& & 0 \leq q_{i} \leq 1, \; i = 1, \ldots, N,\\
&&& \sum_{i=1}^{N} q_{i} \leq M_{d}.
\end{aligned}
\end{equation}

\section{System Design}\label{Section:Section System Design}
\subsection{Successful LoS Reception Probability}
\label{suc_LOS}
An expression for $\slrp_{i}$ can be obtained using fundamentals of stochastic geometry \cite{Haenggi}. That is, from (\ref{scdpLOS}),
\begin{equation*}
    \begin{split}
     \slrp_{i}  
   &=\PR{h_{\hat{x}} > T_{i}r_{\hat{x}}^{\alpha_{L}} \paren{\hat{I}+\hat{N}} \mid r_{\hat{x}} \leq D_{L}} \PR{r_{\hat{x}} \leq D_{L}}\\
      &=\int_{0}^{D_{L}} L_{\hat{I}}\paren{T_{i}x^{\alpha_{L}}}\e{-\hat{N}T_{i}x^{\alpha_{L}}}f_{r_{\hat{x}}}(x) \ dx  ,  
     \end{split}
\end{equation*}
where $\hat{I}=\sum_{y \in \Phi_{d} \setminus \brparen{{\hat{x}}}}h_{y} G^{\prime}r_{y}^{-\alpha_{d}}$
and  $L_{\hat{I}}\paren{S}$ is the Laplace transform of the interference from mmWave D2D links. 
The Laplace transform of the D2D interference is given by 
\begin{equation*}
    \begin{split}
        &L_{\hat{I}}\paren{S}=\ES{\prod_{y \in \Phi_{d} \setminus \{\hat{x}\}}^{}\mathsf{E}_{G^{'},h_{y}}\sqparen{\e{-\frac{G^{\prime}h_{y}r_{y}^{-\alpha_{d}}S}{G_{m}^{2}}}}} \\
        &\stackrel{\text{(a)}}{=}\exp\hspace{-0.1cm}\paren{\hspace{-0.1cm} -\rho \lambda_{u} p_{d}\int_{0}^{2\pi}\hspace{-0.25cm}\int_{0}^{ \infty} \hspace{-0.21cm}\paren{1-\mathsf{E}_{G^{'},h_{y}}\sqparen{e^{-\frac{G^{\prime}h_{y}z^{-\alpha_{d}}S}{G_{m}^{2}}}}}\hspace{-0.1cm}z dz d\phi} \\ 
         &\stackrel{\text{(b)}}{=}\exp\left[-2\pi\rho \lambda_{u} p_{d} \mathsf{E}_{G^{\prime}} \left[\int_{0}^{ D_{L}} \paren{1-\frac{G_{m}^{2}}{1+G^{\prime}z^{-\alpha_{L}}S}}  z dz \right. \right.\\ 
         &\left.\left.+\int_{D_{L}}^{\infty } \paren{1-\frac{G_{m}^{2}}{1+G^{\prime}z^{-\alpha_{N}}S}}z dz \right] \right],
    \end{split}
\end{equation*}
where (a) follows from the probability generating functional (PGFL) of the PPP, $p_d$ is the probability of receiving the requested content via a D2D link, which has to be separately calculated, as shown later Section IV, and the expectation in (b) can be evaluated by using the PDF of $G^{\prime}$ given in Section \ref{Section:System model}, which would result in a product of three exponential functions.  

It is not hard to see that the resulting expression, which has multiple integrals, makes it prohibitively hard for us to use it in a meaningful manner in the optimization problem. We have an $N$-dimensional non-convex constrained optimization problem, and even obtaining the optimum solution numerically is not trivial. Hence, we make few approximations to obtain a mathematically tractable expression for $\slrp_{i}$, $\ie$, we obtain a convex approximation of the objective function such that we can solve the optimization problem in closed form. We note that these approximations are made only to design the content placement policy, and all assumptions are relaxed in the remainder of the paper, which includes the performance evaluation and the numerical evaluations in Sections \ref{section: Performance Analysis} and \ref{section: Numerical results}, respectively. 

Firstly, we neglect small-scale fading with regards to mmWave propagation since it causes only minor changes in received power when the transmitter is within the LoS region \cite{hybridmm&micro,smallCellmmwave}. 
Secondly, we consider the worst case of D2D interference where all the user requests are catered by D2D transmitters, and approximate the random interference using the average worst case interference. To overcome the singularity when computing the D2D interference averaged only over the large-scale fading, we use the bounded path loss model $g(r)= \min\paren{1,r^{-\alpha_{d}}}$ to model the path loss from the interferers, similar to \cite{boundedPathlosslaw}. With these approximations, and by using Campbell's theorem, the average interference can be written as
\begin{multline*}
        \bar{I} = \ES{\sum_{y \in \Phi_{d}\setminus\brparen{x}}G^\prime \min\paren{1,r_{y}^{-\alpha_{d}}}} \\
    \hspace{-2.5cm}    =\frac{\lambda_{u}\rho }{4\pi} \paren{G_{m}\Delta\theta+G_{s}
\paren{2\pi-\Delta\theta}}^{2}\\ 
 \times
\paren{\frac{\alpha_{L}-2D_{L}^{2-\alpha_{L}}}{\alpha_{L}-2}+\frac{2D_{L}^{2-\alpha_{N}}}{\alpha_{N}-2}}.
\end{multline*}
From (\ref{scdpLOS}), and by considering the maximum search discovery distance to initiate a D2D communication link to be $D_{R}$, 
we have
$$
  \slrp_{i} = \PR{r_{\hat{x}} \leq \min\paren{\hat{D}_{i},D_{L}, D_{R}}},
 $$
where 
$$
\hat{D}_{i}= \sqparen{\frac{G_m^2}{T_{i}\bar{I}+T_{i}\hat{N}}}^{\alpha_{L}}.
$$
On the assumption that the transmitters having the $i$-th content stored in their cache form a PPP of intensity $\lambda_{u} \paren{1-\rho} q_{i}$, and by using the distribution of the distance to the nearest MU in a PPP,
we have
\begin{equation}\label{D_ic}
    \slrp_{i}=1-\exp\paren{-\pi\lambda_{u}(1-\rho)q_{i}D_{i,c}^{2}},
\end{equation}
where $D_{i,c}=\min\brparen{\hat{D}_i,D_{L}, D_{R}}$. 

\subsection{Optimum Content Placement}
Once the approximations are applied, it is straightforward to see that the optimization problem becomes convex. The Lagrangian can be written as
\begin{multline*}
\mathcal{L}(\vec{q},\mu)=-1+\sum_{i=1}^{N} \beta_{i} \exp\paren{-\pi\lambda_{u}(1-\rho)q_{i}D_{i,c}^{2}} \\
        +\mu\paren{\sum_{i=1}^{N}q_{i} - M_{d}},    
\end{multline*}
where $\mu$ is the non-negative Lagrangian multiplier. By applying Karush-Kuhn-Tucker (KKT) conditions, we have
$$
q_{i}(\mu)=-\frac{\ln\sqparen{\mu/\paren{\beta_{i}\pi\lambda_{u}(1-\rho)D_{i,c}^{2}} }}{\sqparen{\lambda_{u}(1-\rho)D_{i,c}^{2}}},
$$ 
and according to the first inequality constraint, the optimum $q_i$ should satisfy 
$
    q_{i}^{\star}=\min\brparen{\max\brparen{q_{i}(\mu^{\star}),0},1},
$
and according to the second constraint, which is met with equality, gives us $\sum_{i=1}^{N} \min\brparen{\max\brparen{q_{i}(\mu^{\star}),0},1}= M_{d}.$  This can be used to find $\mu^{\star}$ through a  simple root finding algorithm such as bisection search\cite{whereToCache , hitVSthroughput}.
\vspace{-2mm}
\subsection{User Association}
In a hybrid (multi-tier) wireless network, it is important to associate users with the appropriate tiers to achieve traffic offloading while maintaining QoS constraints. In our system model, an MU may receive content via a LoS D2D link, a NLoS D2D link or a cellular link. Therefore, it is important to recognize the appropriate method of content delivery. From (\ref{D_ic}), one can see that an MU can successfully receive the $i$-th file from a D2D transmitter in the LoS region if the distance to the D2D transmitter is less than $D_{i,c}$.
However, we have assumed the worst case D2D interference when obtaining $D_{i,c}$. 
This means, it may be possible to increase this threshold further without violating the rate constraints, which will facilitate more offloading. Since $\vec{q}$ is now defined, the probability of receiving the requested content from a D2D transmitter within the LoS region can be calculated as
$
    \gamma=\sum_{i=1}^{N} \beta_{i}\paren{1-q_{i}^{\star}}\paren{1-\exp{\paren{-\pi\lambda_{u}(1-\rho)q_{i}^{\star}D_{L}^{2}}}},
$
where we have used the fact that the D2D mode initiates when the required content is not found in self-cache. We can use $\gamma$ to scale the worst-case average interference to have a tighter approximation of the interference from the active D2D transmitters.
Hence, assuming that the interference is dominated by the D2D transmitters in the LoS region, we can obtain an updated distance threshold value that satisfies the QoS requirement as
$$
   \hat{D}_{i,L}= \sqparen{\frac{G_m^2}{T_{i}\gamma\bar{I}+T_{i}\hat{N}}}^{1/\alpha_{L}}.
$$ 
By replacing subscript $L$ with $N$, we can obtain a similar distance threshold $\hat{D}_{i,N}$ for a transmitter in the NLoS region. 

The proposed user association scheme can be summarized as follows. For the $i$-th file, the MU first checks its own cache. If not found, it checks with D2D transmitters who are closer than $$ D_{i,u}= \min\brparen{\hat{D}_{i,L},\max\brparen{\hat{D}_{i,N}, D_L}, D_{R}},$$ for a D2D connection. If both actions fail, the MU connects to the nearest RRH through a cellular link. The process is summarized in Algorithm \ref{uac}.  
\begin{algorithm}[h]
	\caption{User Association Scheme}\label{uac}
	\begin{algorithmic}[1]
	 \STATE  $f_{i} \gets \textnormal{Requested file} $
	 \STATE $A_{L} \gets  \textnormal{circular region with the radius}  \min\brparen{\hat{D}_{i,L},D_R}$
	 \STATE $A_{N} \gets \textnormal{circular region with the radius} \min\brparen{\hat{D}_{i,N},D_R}$
  \IF{ $f_{i}$ \textnormal{in the device self cache} }
    \STATE \textnormal{ Get file from self cache}
  \ELSIF{$\hat{D}_{i,L} \leq D_{L}$ AND $f_{i}$ in $A_{L}$}
    \STATE  \textnormal{ Get file from the closest LoS D2D transmitter}
  \ELSIF{$\hat{D}_{i,N} > D_{L}$ AND $f_{i}$ in $A_{N}$  }
    \STATE \textnormal{Get File from the closest NLoS D2D transmitter}
  \ELSE
    \IF{ $f_i  \textnormal{ in Edge cloud}$ }
        \STATE  \textnormal{ Get file from the edge cloud (fronthaul link)  }
    \ELSE 
        \STATE \textnormal{ Get file from the core network (backhaul link) }
    \ENDIF
  \ENDIF
	    
	\end{algorithmic}
\end{algorithm}

The rationale behind the distance thresholds can be explained as follows. When $\hat{D}_{i,L} < D_L$, it is straightforward that the threshold is $\hat{D}_{i,L}$, as any transmitter outside this (both LoS and NLoS) will not satisfy the rate constraints. When $\hat{D}_{i,L} > D_L$, we particularly focus on transmitters between $D_L$ and $\hat{D}_{i,L}$, who are NLoS according to the channel model. Whether these NLoS transmitters can transmit successfully or not will depend on the value of $\hat{D}_{i,N}$. To this end, if $\hat{D}_{i,N} < D_L$, none of the NLoS transmitters will be able to transmit successfully. Thus, we set the threshold as $D_L$. However, when $D_L \leq \hat{D}_{i,N} \leq \hat{D}_{i,L} $, all NLoS transmitters between $D_L$ and $\hat{D}_{i,N}$ will satisfy the rate constraints, thus we pick $\hat{D}_{i,N}$ as the threshold. Hence, overall, the distance threshold is given by $\max\brparen{\hat{D}_{i,N}, D_{L}}$. The user association policy tries to make use of candidate NLoS transmitters as well, to further facilitate offloading. Note that we will have $\hat{D}_{i,L}> \hat{D}_{i,N}$ for all meaningful link lengths as $\alpha_N > \alpha_L$. 

\section{Performance Analysis}\label{section: Performance Analysis}
Having placed content, and have decided on the user association policy, a performance analysis of the network presented in Section \ref{Section:System model} will be carried out in this section. Note that the assumptions made in Section \ref{Section:Section System Design} are relaxed in this analysis since the assumptions were made only to obtain a mathematically tractable objective function. 

An offloading event occurs when a content request is served by self cache or via a D2D link. To this end, the probability of finding the required content in the device cache itself is given by $ p_{s}=\sum_{i=1}^{N} \beta_{i} q_{i}^{\star}.$ The probability of receiving the requested content via a D2D link is given by 
\begin{equation}\label{p_d}
     p_d= \sum_{i=1}^{N} \beta_{i}\paren{1-q_{i}^{\star}}\paren{1-\exp{\paren{-\pi\lambda_{u}(1-\rho)q_{i}^{\star}D_{i,u}^{2}}}} .
\end{equation}
Thus, $p_d+p_s$ gives us the offloading probability.

We define the successful reception probability ($\spp$) as the probability of an MU receiving content without violating the rate constraints. The $\spp$ through a D2D link is given by 
$
    \sum_{i=1}^{N} \beta_{i}\paren{1-q_{i}^{\star}} \PR{\sinr_{d,i,\hat{x}} \geq T_{i} \cap r_{\hat{x}} \leq D_{i,u}},
$
and similarly, the $\spp$ through the cellular network is given by
$
   \sum_{i=1}^{N} \beta_{i}\paren{1-q_{i}^{\star}}  \e{{-\pi\lambda_{u}(1-\rho)q_{i}^{\star}D_{i,u}^{2}}} \PR{\sinr_{c,i,x} \geq T_{i}}.
$ The sum of these two probabilities and $p_{s}$ gives us $\spp$. 

An expression for $\PR{\sinr_{c,i,x} \geq T_{i}}$ can be obtained using the fundamentals of stochastic geometry, by following a similar approach to the one shown in Section \ref{Section:Section System Design}. 
Considering the closest RRH to be located at $x \in \Phi_R$, 
\begin{equation*}
    \begin{split}
     \PR{\sinr_{c,i,x} \geq T_{i}}
     &=\PR{h_{x} > T_{i}r_{x}^{\alpha_{c}} \paren{\hat{I}_{c}+\hat{N}_{c}}}
     \end{split}
\end{equation*}
\begin{equation*}
=\int_{0}^{\infty} L_{\hat{I}_{c}}\paren{T_{i}z^{\alpha_{c}}}\e{-\hat{N}_{c}T_{i}z^{\alpha_{c}}}f_{r_{x}}(z) \ dz,
\end{equation*}
where $\hat{I}_{c}=\sum_{y \in \Phi_{R} \setminus \brparen{x}}h_{y} r_{y}^{-\alpha_{c}}$, $\hat{N}_{c}=\frac{16 \pi^2N_{o}F_{N}B_{c}}{G_{T}G_{R}P_{c}W_{c}^2}$, and $f_{r_{x}}$ is the PDF of the distance to the nearest RRH, given by $f_{r_{x}}(z)=
    2\pi \lambda_{R} z \e{-\pi \lambda_{R}z^{2} }$, for $ z\geq 0  $. 
Furthermore, we have  
\begin{equation*}
\begin{split}
   L_{\hat{I}_{c}}\paren{S}&= \ES{\prod_{y \in \Phi_{R} \setminus \{x\}}^{}\mathsf{E}_{h_{y}}\sqparen{\e{-h_{y}r_{y}^{-\alpha_{c}}S}}} \\
   &=\exp\paren{-2\pi \lambda_{R}\int_{x}^{\infty} \paren{1-\frac{1}{1+v^{-\alpha_{c}}S} } v dv\ }  .
   \end{split}
\end{equation*}

An expression for $\PR{\sinr_{d,i,\hat{x}} \geq T_{i} \cap r_{\hat{x}} \leq D_{i,u}}$ can be obtained along similar lines, and by appropriately changing the distance limits in the integration. To this end, we get
\begin{equation*}
    \begin{split}
       & \PR{\sinr_{d,i,\hat{x}} \geq T_{i} \cap r_{\hat{x}} \leq D_{i,u}}\\
       &=\int_{0}^{D1}  L_{\hat{I}}\paren{T_{i}z^{\alpha_{L}}} \e{-\hat{N}T_{i}z^{\alpha_{L}}/G^{2}_{m}}f_{r_{\hat{x}}}(z) \ dz \ \\ 
      & +\int_{D_{L}}^{D2}  L_{\hat{I}}\paren{T_{i}z^{\alpha_{N}}} \e{-\hat{N}T_{i}z^{\alpha_{N}}/G^{2}_{m}}f_{r_{\hat{x}}}(x) \ dz \,
    \end{split}
\end{equation*}
where $f_{r_{\hat{x}}}$ is the PDF of the distance to the nearest MU in a PPP of intensity $\lambda_{u}\paren{1-\rho}q_{i}^{\star}$, $D1=\min\brparen{D_{L},\hat{D}_{i,L},D_{R}}$  and $D2=\max\brparen{D_{L},\min\paren{\hat{D}_{i,N},D_{R}}}$. Moreover, 
\begin{equation*}
L_{\hat{I}}\paren{S}\hspace{-0.1cm}=
\exp\hspace{-0.1cm}\paren{\hspace{-0.1cm} -2\pi\rho \lambda_{u} p_{d}\mathsf{E}_{G^{\prime}}\hspace{-0.1cm}\sqparen{
\int_{0}^{\infty} \hspace{-0.2cm}\paren{\hspace{-0.1cm}{1\hspace{-0.1cm}-\hspace{-0.1cm}\frac{G_{m}^{2}}{1+G^{\prime}Sy^{-\alpha(y)}}}\hspace{-0.1cm}}y dy}
},
\end{equation*}
where $\alpha(y)=\hspace{-1mm}\sqparen{1-\hspace{-1mm}\paren{1-\alpha_{L}}\I{y \leq D_L}}\hspace{-1mm}\sqparen{1-\hspace{-1mm}\paren{1-\alpha_{N}}\I{y> D_L}},$ the expectation can be straightforwardly evaluated using the PDF of $G^{\prime}$ given in Section III, and $p_d$ is given by (\ref{p_d}).
 
\vspace{-1mm}\section{Numerical Results and Discussions}\label{section: Numerical results}
In this section, the performance of the proposed scheme is evaluated using simulations. The simulation parameters are set to align with previous works \cite{awareDeviceCaching,smallCellmmwave,whereToCache,journalmmwave} and presented in Table \ref{table_1}. %
We consider $R_i=R$ $\forall$ $i\in \{1,\ldots,N\}$ for simplicity. 
\begin{table}[t]
\caption{Simulation Parameters}
\label{table_1}
\begin{center}
\begin{tabular}{|l||c|}
\hline
 RRH density $\lambda_{R}$ & $10/\textnormal{km}^{2}$\\
\hline
User requesting probability  $\rho $ & 0.5\\
\hline
Path loss exponent ($f_{c}$ = 1 GHz) $\alpha_{c}$ &2.5\\
\hline
Path loss exponent ($f_{c}$ = 28 GHz) $\alpha_{L}$ ,$\alpha_{N}$ &2.1, 4\\
\hline
Transmit power of a RRH $P_{c}$ & 100 mW\\
\hline
Power consumption for backhaul  $P_{b}$ & 1 W\\
\hline
Transmit power of a user device $P_{d}$ & 2 mW \\
\hline
Main lobe gain of the user antenna $G_{m}$ & 9dB\\
\hline
Side lobe gain of the user antenna $G_{s}$ & -9dB\\
\hline
Noise power density $N_{o}$ & -178 dB/Hz\\
\hline
Noise figure $F_{N}$ & 10 dB \\
\hline
Content Library size $N$ & 100\\
\hline
Edge cloud cache capacity $M_{e}$ & 50\\
\hline
Device cache capacity $M_{d}$ & 2\\
\hline
Bandwidth in cellular link $B_{c}$ & 20 MHz\\
\hline
Bandwidth in mmWave link $B_{d}$ & 1 GHz\\
\hline
Maximum search discovery distance of a device $D_{R}$ & 150 m\\
\hline
\end{tabular}
\end{center}
\vspace{-0.5cm}
\end{table}
We compare the performance of the proposed system (S-1) with the system proposed in \cite{whereToCache} (S-2), where the content is placed to maximize the cache hit probability and the content delivery is based on D2D links within a radius of $D_{R}$, which is the maximum discovery distance of an MU. 


\begin{figure}[t]
\centering{\includegraphics[width=0.95\linewidth]{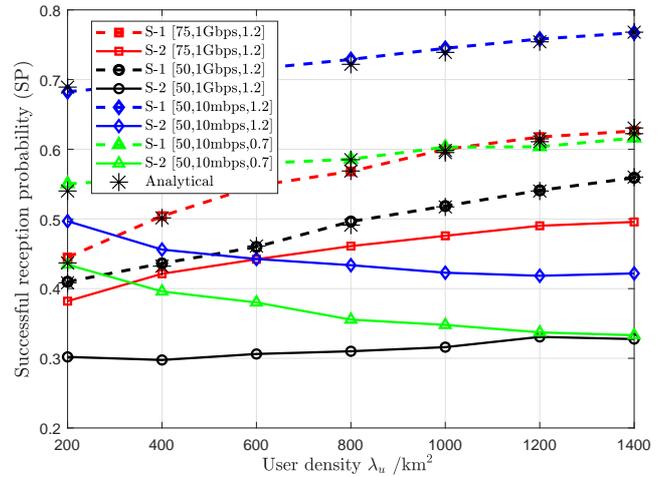}} \vspace{-0.3cm}\caption{The behavior of the overall $\spp$ of the system with $\lambda_u$ for different [$D_{L}$,R,$\epsilon]$ combinations.\vspace*{-2mm}} \label{Fig:ss1}
\end{figure}
Fig. \ref{Fig:ss1} compares the $\spp$ performance of the overall systems, and S-1 outperforms S-2 in all considered scenarios. When $D_{L}$ decreases (blockage density increases), the $\spp$ reduces in both systems. However, we can observe the performance gap between S-1 and S-2 increasing. Since S-1 has prioritized the LoS region for both content placement and delivery, it is more robust to changes in $D_{L}$. On the other hand, S-2, that has focused on $D_{R}$ (which is generally larger than $D_{L}$), may encounter frequent unsuccessful D2D transmissions when the blockage density increases. When both rate constraints and $D_{L}$ reduce, the $\spp$ of S-2 decreases with $\lambda_{u}$. This is due to the increased interference from D2D links and the QoS requirements not being satisfied with S-2. However, since S-1 considers the effect of interference in both the content placement and delivery (user association), the $\spp$ increases with $\lambda_{u}$, making S-1 a promising approach for future ultra-dense networks.
When $\epsilon$ is reduced while keeping the other parameters fixed, the overall success of both the systems reduce. The reduction in $\epsilon$ leads to the content requests spreading out over a large range of files, which in fact reduces the probability of successfully receiving a file over a D2D link. Since both systems have averaged the objective functions over all possible files,  a similar trend is observed for all values of $\epsilon$. 
\begin{figure}[t]
\centering{\includegraphics[width=0.95\linewidth]{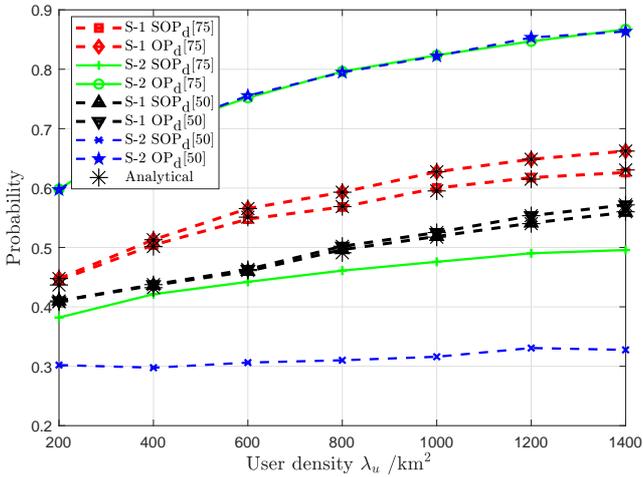}} \vspace{-0.3cm}\caption{Offloading and the successful offloading probabilities of the D2D network for 1 Gbps data rate, where $\epsilon =1.2$.} \label{Fig:so1}
\end{figure}
Fig. \ref{Fig:so1} compares the offloading probability ($\rm{OP}_d$) of the two systems. S-2 having a higher offloading probability is rather obvious since considering $D_R$ leads to a larger offloading region with S-2. However, the figure also conveys that a considerable portion of the offloaded traffic in S-2 will not be successfully delivered, and hence, the successful offloading probability ($\rm{SOP}_{d}$) of S-2 is lower than S-1. When $D_{L}$ reduces, the unsuccessful offloading of S-2 increases, 
but in S-1, the gap between the two offloading probabilities remains low as it is more robust to changes in $D_{L}$, as described with respect to Fig. \ref{Fig:ss1}. 


\begin{figure}[t]
\centering{\includegraphics[width=0.95\linewidth]{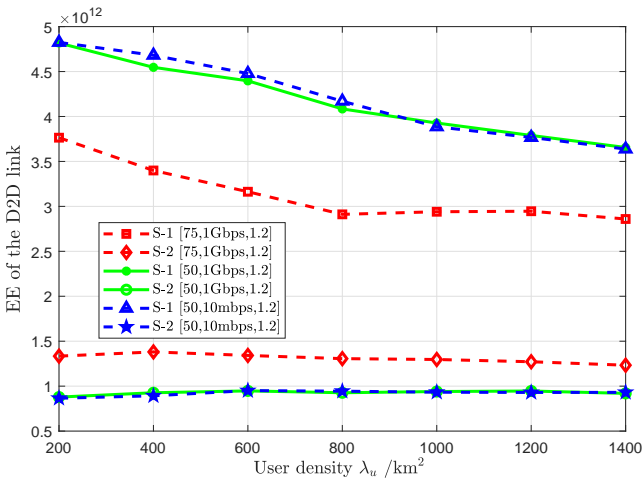}} \vspace{-0.3cm}\caption{The Energy Efficiency of the D2D link with $\lambda_u$ for different [$D_{L}$,R,$\epsilon]$ combinations.\vspace*{-2mm}} \label{Fig:energy}
\end{figure}

Energy efficiency (EE) can be computed by the ratio between the average throughput from successful transmissions and the average power consumption, per user request. Some insights on the EE of the two systems can be inferred from Fig. \ref{Fig:so1}. 
When $\lambda_{u}$ increases, the probability of initiating a D2D link increases in both schemes. This leads to the average throughput from successful transmissions increasing, and the average power consumption decreasing. However, the average throughput from successful transmissions of S-1 increases at a higher rate with $\lambda_{u}$ compared to S-2. On the other hand, the average power consumption of S-2 decreases at a higher rate with $\lambda_{u}$ compared to S-1 due to the higher offloading. Therefore, the overall EE of the two systems are almost similar. However, one can identify that the EE of D2D links is significantly higher for S-1, compared to S-2, which is illustrated in Fig. \ref{Fig:energy}. This is due to the unsuccessful D2D transmissions in S-2 resulting in device energy wastage. When $\lambda_{u}$ is varied from 200 to 1400 /$\textnormal{km}^{2}$, for $D_{L}=75$m and $\epsilon =1.2$, on average, a 1.3 fold improvement in terms of EE is observed with S-1 compared to S-2. This improves significantly when $D_{L}$ is further reduced. For example, on average, a 3.5 fold improvement can be observed at $D_{L}=50$m. This depicts the superior performance of S-1 in ultra-dense networks with high blockage.  

\section{Conclusions}\label{section: conclusion}
 The performance of a cache enabled D2D network, where D2D communications occur exclusively in the mmWave band, has been studied using a stochastic geometric framework. As a result, a novel content caching scheme in user devices to maximize the successful content delivery probability of LoS D2D links has been introduced. The performance gains of the proposed content placement and delivery schemes have been highlighted through simulations. The numerical results have shown that the proposed scheme achieves higher successful content offloading, improved energy efficiency while satisfying QoS requirements of the users, and superior performance in ultra-dense networks with high blockage.

\ifCLASSOPTIONcaptionsoff
  \newpage
\fi
\bibliographystyle{ieeetr}

\begin{thebibliography}{10}

\bibitem{D2D_survey2018}
F.~{Jameel}, Z.~{Hamid}, F.~{Jabeen}, S.~{Zeadally}, and M.~A. {Javed}, ``A
  survey of device-to-device communications: Research issues and challenges,''
  {\em IEEE Commun. Surv. Tut.}, vol.~20, pp.~2133--2168, Apr. 2018.

\bibitem{5GD2D_2018}
R.~I. {Ansari}, C.~{Chrysostomou}, S.~A. {Hassan}, M.~{Guizani}, S.~{Mumtaz},
  J.~{Rodriguez}, and J.~J. P.~C. {Rodrigues}, ``5{G} {D2D} networks:
  Techniques, challenges, and future prospects,'' {\em IEEE Syst. J.}, vol.~12,
  pp.~3970--3984, Dec. 2018.

\bibitem{whereToCache}
Z.~Chen and M.~Kountouris, ``{D2D} caching vs. small cell caching: Where to
  cache content in a wireless network?,'' in {\em Proc. IEEE International
  Workshop on Signal Processing Advances in Wireless Communications}, pp.~1--6,
  Jul. 2016.

\bibitem{hitVSthroughput}
Z.~Chen, N.~Pappas, and M.~Kountouris, ``Probabilistic caching in wireless {
  D2D} networks: Cache hit optimal versus throughput optimal,'' {\em IEEE
  Commun. Letters}, vol.~21, pp.~584--587, Mar. 2017.

\bibitem{awareDeviceCaching}
N.~Giatsoglou, K.~Ntontin, E.~Kartsakli, A.~Antonopoulos, and C.~Verikoukis,
  ``{D2D}-aware device caching in mm{W}ave-cellular networks,'' {\em IEEE J.
  Sel. Areas Commun.}, vol.~35, pp.~2025--2037, Sep. 2017.

\bibitem{adhoc-mmwave}
S.~Vuppala, T.~X. Vu, S.~Gautam, S.~Chatzinotas, and B.~Ottersten,
  ``Cache-aided millimeter wave ad-hoc networks,'' in {\em Proc. IEEE Wireless
  Communications and Networking Conference}, pp.~1--6, Apr. 2018.

\bibitem{coverageRateLosBall}
T.~Bai and R.~W. Heath, ``Coverage and rate analysis for millimeter-wave
  cellular networks,'' {\em IEEE Trans. Wireless Commun.}, vol.~14,
  pp.~1100--1114, Feb. 2015.

\bibitem{losBall}
Y.~Zhu, L.~Wang, K.~Wong, and R.~W. Heath, ``Secure communications in
  millimeter wave ad hoc networks,'' {\em IEEE Trans. Wireless Commun.},
  vol.~16, pp.~3205--3217, May 2017.

\bibitem{journalmmwave}
J.~G. Andrews, T.~Bai, M.~N. Kulkarni, A.~Alkhateeb, A.~K. Gupta, and R.~W.
  Heath, ``Modeling and analyzing millimeter wave cellular systems,'' {\em IEEE
  Trans. Commun.}, vol.~65, pp.~403--430, Jan. 2017.

\bibitem{Haenggi}
M.~Haenggi, J.~G. Andrews, F.~Baccelli, O.~Dousse, and M.~Franceschetti,
  ``Stochastic geometry and random graphs for the analysis and design of
  wireless networks,'' {\em IEEE J. Sel. Areas Commun.}, vol.~27,
  pp.~1029--1046, Sep. 2009.

\bibitem{hybridmm&micro}
F.~Wang, H.~Wang, H.~Feng, and X.~Xu, ``A hybrid communication model of
  millimeter wave and microwave in {D2D} network,'' in {\em Proc. IEEE
  Vehicular Technology Conference}, pp.~1--5, May 2016.

\bibitem{smallCellmmwave}
Y.~Zhu, G.~Zheng, L.~Wang, K.~Wong, and L.~Zhao, ``Performance analysis and
  optimization of cache-enabled small cell networks,'' in {\em Proc. IEEE
  Global Telecommunications Conference}, pp.~1--6, Dec. 2017.

\bibitem{boundedPathlosslaw}
N.~Deng, M.~Haenggi, and Y.~Sun, ``Millimeter-wave device-to-device networks
  with heterogeneous antenna arrays,'' {\em IEEE Trans. Commun.}, vol.~66,
  pp.~4271--4285, Sep. 2018.

\end{thebibliography}




\end{document}